\documentclass[pra,twocolumn]{revtex4}
\usepackage{graphicx}
\usepackage{dcolumn}
\usepackage{bm}
\usepackage{amsmath}
\usepackage{amsfonts}
\usepackage{psfrag}
\usepackage{graphicx}
\usepackage{color}

\newcommand{\Eq}[1]{Eq. (\ref{#1})}

\begin{document} 

\title{Comment on ``System-environment coupling derived by Maxwell's boundary conditions from the weak to the ultrastrong light-matter coupling regime''}
\author{Simone \surname{De Liberato}}
\affiliation{School of Physics and Astronomy, University of Southampton, Southampton, SO17 1BJ, United Kingdom}

\begin{abstract}
In a recent work \cite{Bamba13}, Bamba and Ogawa developed a microscopic model describing the field of a photonic cavity coupled to a matter exciton-like resonance. One of the results they obtain studying such a model is that, in the ultrastrong coupling regime, usually safe approaches can give wrong results for the dissipation rates of the polaritonic excitations. In particular the dissipation rates calculated applying the rotating wave approximation on the system-environment coupling qualitatively differ from the ones calculated using a microscopic theory based on the quantum electrodynamics for dielectric media. Here I show that this result is an artifact, caused by an inconsistent application of the rotating wave approximation and by a questionable parameter choice. 
\end{abstract}

\maketitle

\begin{figure*}[t]
\begin{center}
\includegraphics[width=5.5cm]{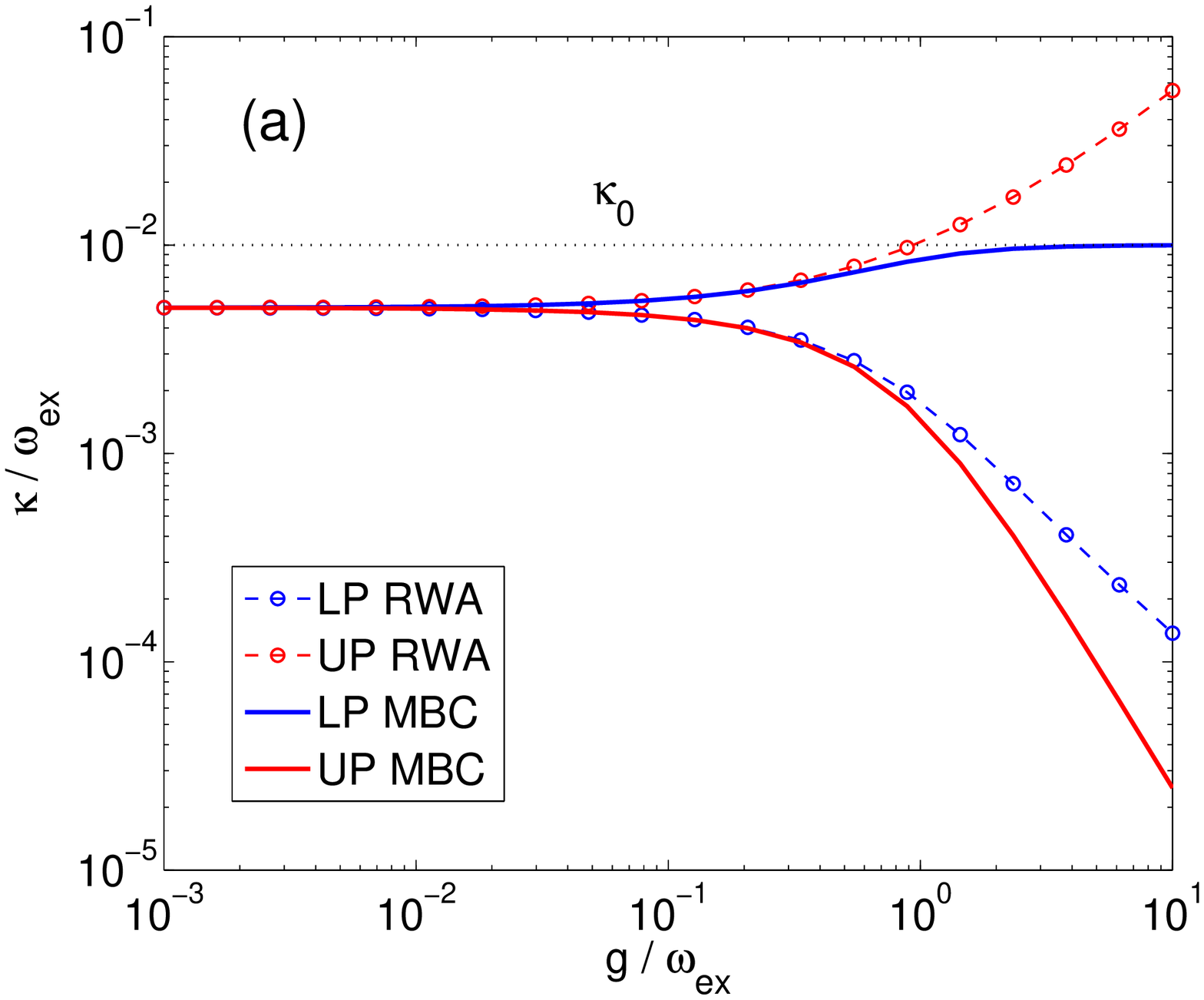}
\includegraphics[width=5.5cm]{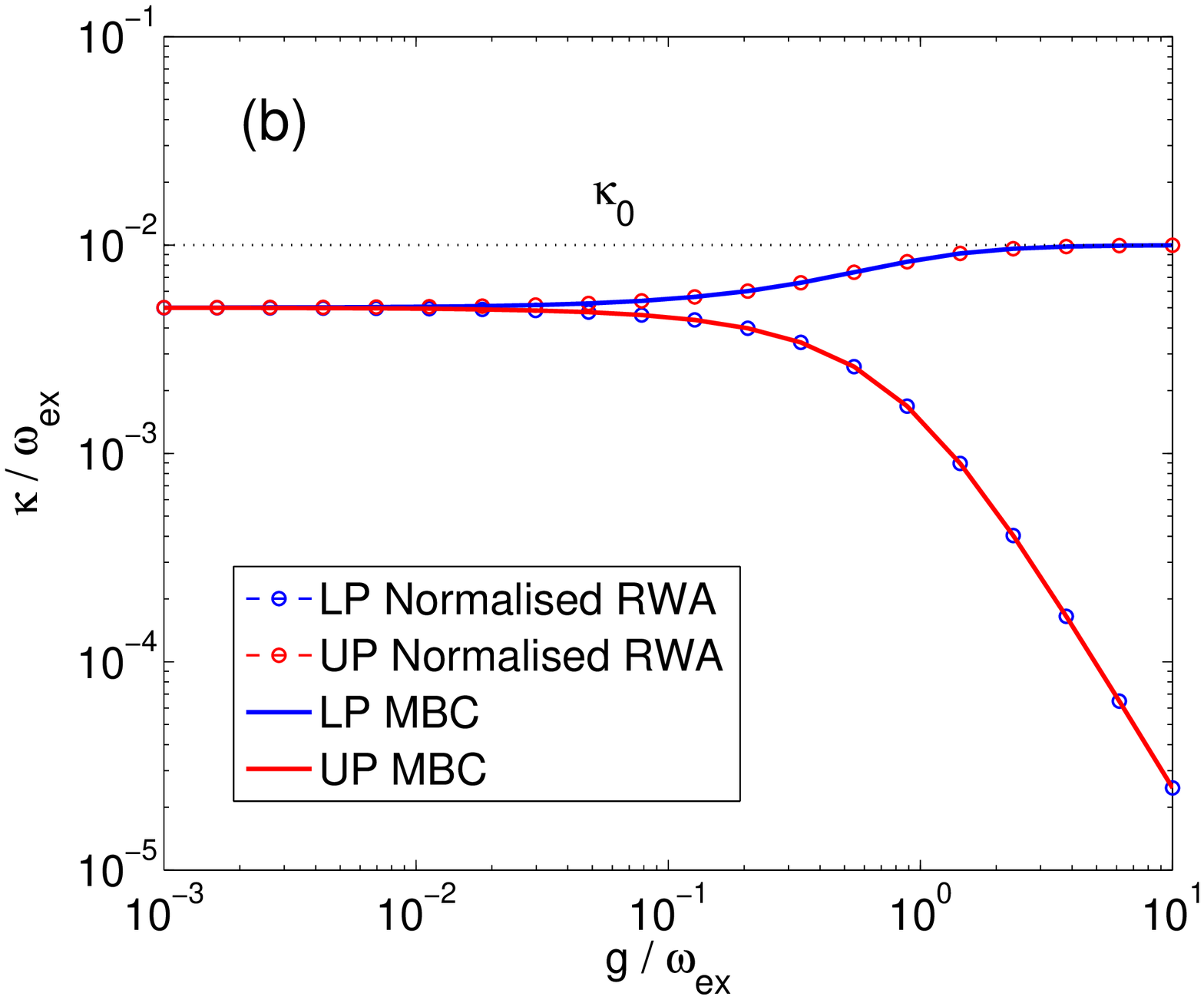}
\includegraphics[width=5.5cm]{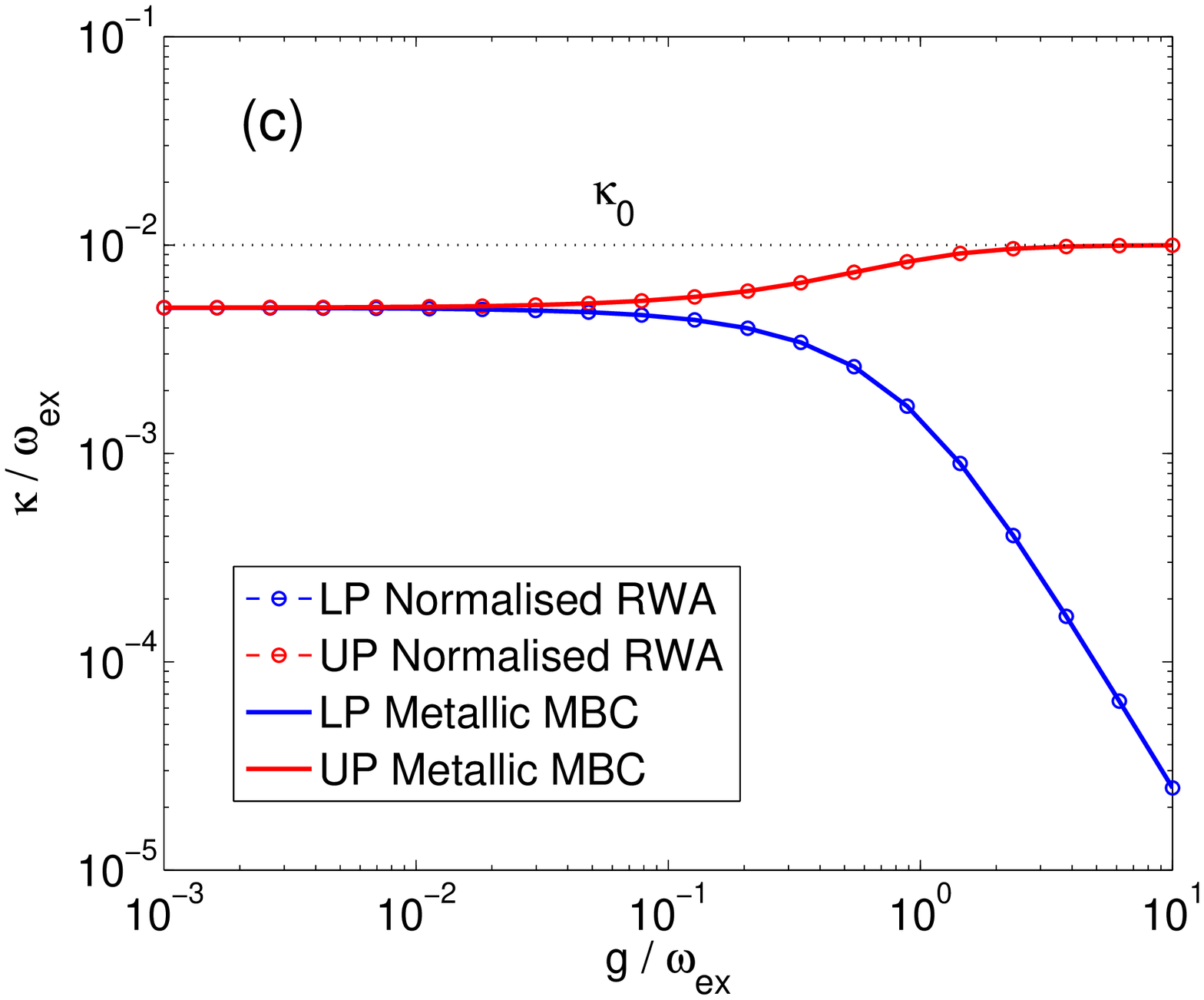}
\caption{\label{F1} In panel (a) there is a 
comparison of the dissipation rates obtained in Ref. \cite{Bamba13} using the RWA  and the MBC as a function of the vacuum Rabi frequency $g$.
In panel (b) the normalised RWA in \Eq{RWAresmod} is used instead that the formula in \Eq{RWAres}.
In panel (c) the results for the normalised RWA are compared to the ones obtained from \Eq{kt} for metallic MBC.
 In this figure the frequency of the first cavity mode is equal to $\omega_{\text{ex}}$ and  $\kappa_0$ is its dissipation rate.}
\end{center}
\end{figure*}
In the light-matter ultrastrong coupling (USC) regime many usually safe approximations fail and it is thus necessary to pay great attention when studying this non-perturbative interaction regime. In particular, neglecting the light-matter interaction while calculating the system- environment coupling in the USC regime can lead to wrong or unphysical results
\cite{DeLiberato09,Beaudoin11,Bamba12}.

A simple way to study open quantum systems in the USC regime is to apply the rotating wave approximation (RWA) on the system-environment coupling in the dressed states basis. In this way positive and negative frequency excitations do not mix and the global ground state is the product of the systemÕs and environmentÕs ground states \cite{Ciuti06, Bamba12}.
Until now no work had addressed the important question of the reliability and generality of such an approach.

In Ref. \cite{Bamba13} the authors developed a microscopic model describing a photonic cavity filled with infinite-mass oscillators. Using Maxwell boundary conditions (MBC) they calculated the lifetimes of the cavity eigenmodes due to the finite reflectivity of one of the mirrors, as a function of the strength of the light-matter coupling. The microscopic calculation proves that the lack of positive and negative frequency mixing is not a simplification imposed by the RWA but it emerges naturally when one considers in a consistent way a microscopic theory based on the quantum electrodynamics for dielectric media. One of the consequences of such a theory is that in the USC regime the dissipation rates calculated using MBC differ qualitatively from the ones obtained using the usual system-environment RWA approach. This result is very surprising and, if true, it could have a major impact on future investigations on USC physics, as it would imply that in this regime the system observables can critically depend on microscopic details. The bottom panel of Fig. 2 in Ref. \cite{Bamba13}, replicated here in the panel (a) of Fig. 1, shows the dissipation rates for a resonant cavity mode, calculated using the MBC approach compared with the ones calculated using the RWA. They clearly differ in two main regards: (I) the values given by the RWA are up to one order of magnitude larger that the MBC ones and (II) the MBC gives larger losses for the lower polariton, while the RWA for the upper one.
 
In this comment I will show that such a discrepancy is an artifact. Point (I) is the consequence of an inconsistent application of the RWA while point (II) is the consequence of a questionable parameter choice. If such mistakes are corrected the two approaches give fully consistent results. 

About point (I), in Ref. \cite{Bamba13} the following formula is used for the dissipation rate  of the $m^{\text{th}}$ polaritonic mode in the RWA
\begin{eqnarray}
\label{RWAres}
\kappa^{\text{RWA}}_{jm}=\lvert \omega_{jm} \lvert^2 \kappa_m,  
\end{eqnarray}
where $j=\text{L},\text{U}$, indexes the two polaritonic branches. This result is found starting
from the system-environment coupling Hamiltonian
\begin{eqnarray}
\label{HSR}
H_{\text{S-E}}&=&\sum_m \int d\omega\, i\hbar\sqrt{\frac{\kappa_m}{2\pi}}\lbrack \alpha(\omega)^{\dagger}a_m-a_m^{\dagger}\alpha(\omega) \rbrack,
\end{eqnarray}
and expressing the cavity photon operators as a function of the polaritonic operators 
\begin{eqnarray}
\label{Pola}
a_m&=&\sum_{j=\text{L},\text{U}} (\omega^*_{jm}p_{jm}-y_{jm}p_{jm}^{\dagger}).
\end{eqnarray}
Neglecting the resulting anti-resonant terms, that is considering $y_{jm}=0$,  \Eq{RWAres} is obtained.
The problem with this procedure is that, in order to respect bosonic commutation relations, the coefficients in \Eq{Pola} have to respect the normalisation condition
 \begin{eqnarray}
\label{Norm}
\sum_{j=\text{L},\text{U}} \lvert \omega_{jm}\lvert^2-  \lvert y_{jm}\lvert^2 &=&1.
\end{eqnarray}
To simply neglect the $y_{jm}$ coefficients amounts to consider non-normalised, and thus non-bosonic polaritonic operators.
The problem can be solved renormalising the Hopfield coefficients after having put the antiresonant terms to zero. This gives the dissipation rate
\begin{eqnarray}
\label{RWAresmod}
\tilde{\kappa}^{\text{RWA}}_{jm}=\frac{\lvert \omega_{jm} \lvert^2}{\sum_{j=\text{L},\text{U}} \lvert \omega_{jm}\lvert^2} \kappa_m.  
\end{eqnarray}
In Fig. 1 (b) there is a comparison between the MBC results from Ref. \cite{Bamba13} and  the results obtained by the normalised RWA in \Eq{RWAresmod}. As we can see problem (I) has been completely solved. The dissipation rates obtained using the two approaches are consistent, although still inverted.

About point (II) Bamba and Ogawa consider for simplicity their parameter $\Lambda(\omega)$ to be frequency independent over the frequency range of interest. As $\Lambda(\omega)=\eta(\omega)\omega / c$ this implies that the 
mirror's permittitivity $\eta(\omega)$ is proportional to $\omega^{-1}$. This seems a physically unjustified assumption especially given that, as the mirror in their model is infinitely thin, it would seem more natural to assume it is metallic, having thus a permittitivity roughly proportional to $\omega^{-2}$.
Taking into account this different frequency dependency, the Eq. (26) of Ref. \cite{Bamba13} 
\begin{eqnarray}
\label{k}
\kappa_{\text{MBC}}(\Omega)\simeq \frac{\kappa_0}{1+(\Omega/\omega_{ex})^2},
\end{eqnarray}
giving the loss rates using the MBC approach becomes instead
\begin{eqnarray}
\label{kt}
\tilde{\kappa}_{\text{MBC}}(\Omega)\simeq \frac{\kappa_0}{1+(\omega_{ex}/\Omega)^2}.
\end{eqnarray}
In Fig. 1 (c) we can see that the loss rates calculated using MBC with metallic mirrors from \Eq{kt} and the ones using the normalised RWA formula from \Eq{RWAresmod} 
are in very good agreement, and that also the problem highlighted in point (II) has thus disappeared.

The problem of understanding the reliability of the RWA phenomenological approach to system-environment coupling in the USC regime remains open. The very good agreement between the results obtained using the metallic MBC in \Eq{kt} 
and the RWA with frequency independent loss rates in \Eq{RWAresmod} is not well understood, and it is at this point unclear if some form of generality does hold, or if in the USC regime a degree of microscopic modeling is necessary to obtain quantitative results.
Still this comment shows that the model developed in Ref. \cite{Bamba13} does not disprove the applicability of the RWA approach but, on the contrary, it gives a first microscopic justification of its validity, at the very least for the particular system considered.

I wish to thank M. Bamba and N. Shammah for fruitful discussions and to acknowledge the support of the European Commission under the Marie Curie IEF Program, project BiGExPo.

\end{document}